\documentclass[osajnl,twocolumn,showpacs,10pt]{revtex4-1} 
\usepackage{amsmath,amssymb,graphicx}
\usepackage{epsfig}

\begin{document}
\title{Optimizing a spin-flip Zeeman slower}
\author{L. Zhao}
\email{lichao@okstate.edu}
\author{J. Jiang}
\author{Y. Liu}
\affiliation{Department of Physics, Oklahoma State University,
Stillwater, OK 74078}

\begin{abstract}
We present a design of a spin-flip Zeeman slower controlled by a
fast feedback circuit for a sodium Bose-Einstein condensate
apparatus. We also demonstrate
how the efficiency of the slower strongly depends on its
intrinsic parameters, and compare these observations with a few
theoretical models. Our findings lead to a simple three-step
procedure of designing an optimal Zeeman slower for neutral atoms,
especially for those atomic species with high initial velocities,
such as sodium and lithium atoms.
\end{abstract}


\maketitle
\section{Introduction}
Laser cooling and trapping neutral atoms with a magneto-optical
trap (MOT) has become an important step in a successful production
of ultracold quantum gases~\cite{mot}. To improve the capture
efficiency of a MOT, a number of slowers were invented to slow hot
atoms before they overlap with the MOT~\cite{ketterle,
mot,durfee,phillips,naik,gibble}. Atoms and a resonant laser beam
counter propagate in a slower. The longitudinal velocity and
corresponding Doppler shift of these atoms decrease after they
absorb resonant photons. After a few such absorptions, these
slowed atoms are no longer resonant with the laser beam and thus
cannot be further slowed down. In order to continuously slow atoms
along the laser beam path, one can vary the laser frequency
accordingly as with the frequency chirp method~\cite{Wallis} or by
using broadband lasers~\cite{zhu}. Another widely-used method is
to compensate differences in the Doppler shift with a spatially
varying magnetic field generated by a Zeeman slower while keeping
the laser frequency
unchanged~\cite{mot,durfee,phillips,Stamper-Kurn2010,cheiney,bell,naik,
slower2004, Meschede1999}. In this paper, we present the design
and construction of a spin-flip Zeeman slower controlled by a fast
feedback circuit for a sodium Bose-Einstein condensate (BEC)
apparatus. An efficient method of optimizing a slower with our
simulation program and by monitoring the number of atoms trapped
in the MOT is also explained. In addition, our data demonstrates how the efficiency of a slower
strongly depends on a few of its intrinsic parameters, such as the
intensity of the slowing laser beam and the length of each section
in the slower. These findings result in a simple three-step
procedure of designing an optimal Zeeman slower for neutral atoms,
especially for those atomic species with high initial velocities,
for example lithium atoms.
\section{Experimental Setup}
\begin{figure}[t]
\includegraphics[width=85mm]{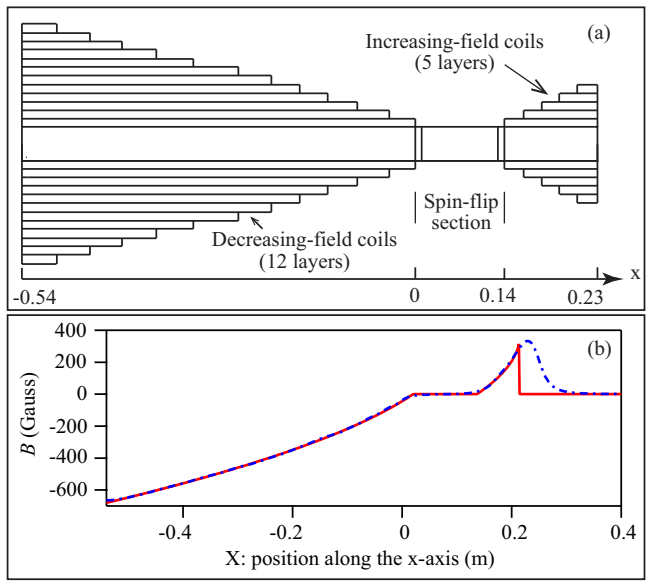}
\caption{(color online) (a)
schematic of our spin-flip Zeeman slower. The first
layer of the decreasing-field coils has 188 turns and its
length is 0.54~m. The second layer is 0.51~m long and wrapped on
the surface of the first layer. Similarly, the following layers
are wrapped on the surface of its corresponding previous layer.
The increasing-field section has five layers and is constructed in a
similar way. All axes are not to
scale. (b) A comparison between a theoretical
prediction (solid red line) and the performance of our optimized
Zeeman slower (dash-dotted blue line), with $\delta=-512$~MHz and
$\eta$ in decreasing- and increasing-field sections being set at 0.65. Atoms propagate along the
positive $x$-axis direction. The MOT center locates at 0.45 m, where
residual magnetic fields created by the slower are negligible.} \label{theory}
\end{figure}

A sodium beam is created by an oven consisting of a half nipple
and an elbow flange. A double-sided flange with a 6~mm diameter
hole in the center is used to collimate the atomic beam. To
collect scattered atoms, a cold plate with a 9~mm diameter center
hole is placed before a spin-flip Zeeman slower and kept at 10
$^\circ$C with a Peltier cooler. Our multi-layer slower has three different
sections along the $x$ axis (i.e., a decreasing-field section, a
spin-flip section, and an increasing-field section), as shown in
Fig.~\ref{theory}(a). Compared with
the single-layer Zeeman slower with variable pitch
coils~\cite{bell}, this multi-layer design provides us enormous
flexibilities in creating large enough $B$ for slowing atoms with
high initial velocities (e.g., sodium and lithium atoms). The first section of our Zeeman slower produces a magnetic field
with decreasing magnetic field strength $B$. We choose $B \sim650$
Gauss at the entrance of the slower, so the slowing beam only
needs to be red-detuned by $\delta$ of a few hundred MHz from the
D2 line of $^{23}$Na atoms. This frequency detuning is easily
achieved with an acousto-optic modulator, but is still large
enough to avoid perturbing the MOT. The spin-flip section
is simply a bellow as to maintain $B=0$ for atoms to be fully
re-polarized and also to damp out mechanical vibrations generated
by vacuum pumps. The increasing-field section creates a magnetic
field with increasing $B$ but in the opposite direction to that of
the decreasing-field section, which ensures the magnetic field
quickly dies off outside the slower. This slower can thus be placed
close to the MOT, which results in more atoms being captured. The
MOT setup is similar to that of our previous work~\cite{JiangBEC}
and its maximum capture velocity $v_{\rm{c}}$ is $\sim$~50~m/s.

To precisely adjust $B$ inside the slower, all layers of magnetic coils are divided into four groups, and different layers in
each group are connected in series and controlled by one fast
feedback circuit. A standard ring-down circuit consisting of a
resistor and a diode is also connected in parallel with each coil
for safely shutting off the inductive current in the coil.
An important chip in our control circuit is a high power metal
oxide semiconductor field effect transistor (MOSFET) operated in the linear mode. We use a number of MOSFETs
connected in parallel and mount them on the top of a water-cooled
cold plate to efficiently cool them. A carefully chosen resistor
R$_s$ of 50~m$\Omega$ is connected to each MOSFET's source
terminal to encourage equal current splitting among the MOSFETs in
parallel. R$_s$ also limits MOSFET's transconductance to a
narrow range, which enables our feedback circuit to control both
low and high electric currents with a single set of gains. The
design of this feedback circuit is available upon request.

\section{Optimization}
When neutral atoms pass through the Zeeman slower, only those
atoms with a longitudinal velocity $v(x) = -[2\pi \delta+\mu
B(x)/\hbar]/k$ are resonant with the slowing beam and can be
slowed. Here $\mu$ is the magnetic moment, $\hbar$ is the reduced
Planck's constant, $k$ and $\delta$ are the wavevector and
frequency detuning of the laser beam, respectively. The actual
deceleration $a_s=\eta a_{\rm{max}}$ in the slower is thus given
by
\begin{equation}
\frac{dB(x)}{dx}=-a_s \frac{\hbar k}{\mu v}=\eta
a_{\rm{max}}\frac{\hbar k^2}{\mu [2\pi \delta+\mu B(x)/\hbar]}~,
\label{1}
\end{equation}
where $\eta$ is a safe factor to account for magnetic field
imperfections in a given slower and the finite intensity of the
slowing laser beam, and $a_{\rm{max}}=\hbar k\Gamma / 2m$ is the
maximal achievable deceleration. $m$ and $\Gamma$ are the mass and
the natural linewidth of the atoms, respectively.

Our largest MOT is achieved when we match $B(x)$ inside
the slower as precisely as possible to a prediction derived from Eq.~\ref{1} with $\eta$
being set at 0.65 in decreasing- and increasing-field sections and $v_{f}=$~40 m/s, as shown in
Fig.~\ref{theory}(b). Here $v_{f}$ is the velocity of atoms at the
end of the slower. $N$, the number of atoms in a MOT, can also be
boosted by a larger atomic flux with a hotter atomic oven.
However, this is generally not a favorable method due to two
reasons. First, a hotter oven generates atoms with higher initial
average velocities, but a slower can only handle entering atoms of
a certain maximum velocity. Second, alkali metals' consumption
rates sharply increase with the oven temperature.

We find that convenient parameters to adjust are slowing beam's
intensity $I$ and frequency detuning $\delta$, electric current
$i$ in each magnetic coil, and the length of each section of the
slower. These parameters, however, cannot be optimized
independently since there is a strong correlation among them. In
order to efficiently optimize the slower, we developed a computer
program to simulate $B(x)$ and compared the simulation results to
the actual magnetic field strengths in the slower under many
different conditions (e.g., at various values of $i$). The
actual magnetic field strengths were measured with a precise Hall
probe before the slower was connected to the vacuum apparatus. The
agreements between the simulation results and the measurements are
good, i.e., the discrepancies are $<$~5 \%. This simulation
program can thus mimic the actual performance of a
Zeeman slower and provide a reasonable tuning range for each aforementioned
parameter, which allows for efficiently optimizing the slower. Our simulation program is available upon request.

One common way to evaluate a slower's performance is from knowing
the exact number of atoms whose velocities are smaller than
$v_{\rm{c}}$ with a costly resonant laser. In the next four
subsections, we show that a convenient
detection method of optimizing a slower is to monitor the number
of atoms captured in the MOT, i.e., more atoms trapped in a given
MOT setup indicate a better performance of the slower.

\begin{figure}[t]
\includegraphics[width=85mm]{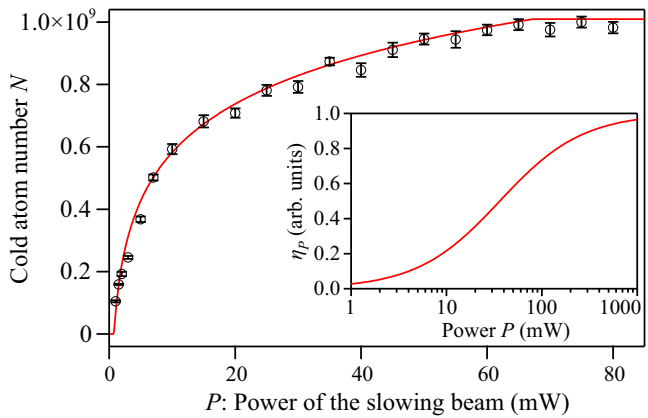}
\caption{(color online) $N$ as a function of the
slowing beam power $P$ with 1~mW corresponding to $I_0/I_s=2.5$. The
solid line is a fit based on Eq.~\ref{Npower} with $\eta$ in decreasing- and increasing-field
sections being set at 0.65. Inset: $\eta_p$ as a function of $P$ for our apparatus.} \label{power}
\end{figure}

\subsection{Intensity of the slowing beam}
Figure~\ref{power} shows that the MOT capture efficiency strongly
depends on the slowing beam power $P$, i.e., $N$ increases with $P$ and then stays at
its peak value $N_{\rm max}$ when $P$ is higher than a critical value. This can be understood from the
relationship between the safe factor $\eta$ of a slower and the
slowing beam intensity $I$. Based on Ref.~\cite{Meschede1999}, the safe factor $\eta_{\rm laser}$ at
a finite $I$ can be expressed as,
\begin{equation}
\eta_{\rm laser}=\frac{I/I_s}{1+I/I_s+[\frac{2 \pi \delta+\mu
B/\hbar+kv}{\Gamma/2}]^2} \leq \eta_{\rm laser}^{\rm max} = \frac{I/I_s}{1+I/I_s}~,\label{safefactor1}
\end{equation}
where $I_s$ is the saturation intensity of neutral atoms, e.g., it is $6.26~\rm{mW/cm^2}$ for
sodium atoms. 
Eq.~\ref{safefactor1} implies that the safe factor of any optimal Zeeman slower has
an upper limit, $\eta_{\rm laser}^{\rm max}$, as long as the slowing beam
intensity is fixed. In other words, a bigger $\eta$ in the
decreasing- or increasing-field section does not always lead to a more
efficient Zeeman slower if $I$ is given.

For a Gaussian slowing beam with a width $W$, its intensity $I$
can be expressed as $I(r)=I_0\cdot e^{-\frac{2r^2}{W^2}}$. Here
$r$ is the distance away from the slowing beam center and $I_0=2
P/(\pi W^2)$ is the beam intensity at $r=0$. $N$ can be given by,
\begin{align}
N(P) &=\frac{N_{\rm max}}{\pi R_0^2}
\int_0^{R_0}H\left[\eta_{\rm laser}^{\rm max}(r)-\eta\right]2 \pi rdr \nonumber \\
&=\frac{N_{\rm max}}{R_0^2/2}\int_0^{R_0}H\left[\frac{
P\cdot \exp(-\frac{2r^2}{W^2})}{\frac{\pi}{2} W^2 I_s+
P\cdot \exp(-\frac{2r^2}{W^2})}-\eta\right]rdr~.\label{Npower}
\end{align}
Here $H[r]$ is a unit step function of $r$ to account for the
fact that atoms can be efficiently slowed only when
$\eta~\leq~\eta_{\rm laser}^{\rm max}(r)$, and $R_0$ is the radius of the
atomic beam. Figure~\ref{power} shows that our data taken with $\eta=0.65$ in decreasing- and increasing-field sections can be well fitted by Eq.~\ref{Npower} and $N$ saturates at $P \geq 70$~mW. This indicates that 70~mW is the minimum power to achieve $\eta=0.65$ for our apparatus. We can thus define a $\eta_p$, the preferred $\eta$ in decreasing- and increasing-field sections at a given $P$, and derive its expression from Eq.~\ref{Npower} as follows,
\begin{equation}
\eta_p=P\cdot \exp(-\frac{2R_0^2}{W^2})/[\frac{\pi}{2} W^2 I_s+
P\cdot \exp(-\frac{2R_0^2}{W^2})]~. \label{etap}
\end{equation}
 
The predicted $\eta_p$ as a function of $P$ for our apparatus is shown in the inset in Fig.~\ref{power}, which implies $P$ sharply increases with $\eta_p$ and is infinitely large at
$\eta_p=1$. Therefore,
the first step in designing an optimal Zeeman slower is to
determine $\eta_p$ from Eq.~\ref{etap} based on the
available slowing beam power.
\begin{figure*}[t]
\includegraphics[width=175mm]{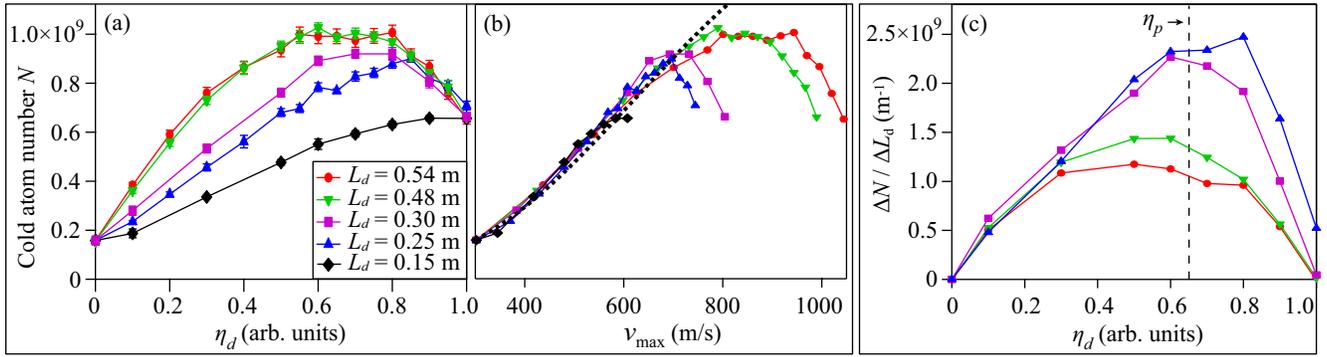}
\caption{(color online) $N$ as a
function of $\eta_d$ (a) and $v_{\rm max}$ (b) with $\eta$ being
set at 0.65 in the increasing-field section, $\delta=-512$~MHz, and
$P=70$~mW. Dotted black line in Panel
(b) is a fit based on Eqs.~\ref{Qs}-\ref{vMax}. (c) $\Delta N/\Delta L_d$ as a function
of $\eta_d$. Here $\Delta L_d=L_d-0.15$~m, and $\Delta N$ is the
number of extra atoms being slowed when $L_d$ is set at a value
longer than 0.15~m. Black dashed line marks the position of
$\eta_p$, which is determined by the slowing
laser beam power.} \label{decreasing}
\end{figure*}

\subsection{The decreasing-field section}
To focus on the
decreasing-field section, our data shown in this subsection are
taken with $\eta$ being set at 0.65 in the increasing-field section, a
fixed distance between the atomic oven and the MOT center
to maintain a fixed solid angle for an atomic beam,
$\delta=-512$~MHz, and $P=70$~mW which implies
$\eta_p$ is 0.65. Based on the discussions in
Refs.~\cite{durfee, Stamper-Kurn2010}, $N$ can be expressed as
\begin{align}
N &= \int_0^{v_{\rm{max}}} N_0(v) f(v) dv~, \label{Qs}
\end{align}
where $N_0(v)\propto v^3 e^{-\frac{m v^2}{k_B T}}$ is the initial
number of atoms created by the oven, $k_B$ is the Boltzmann
constant, and the oven temperature $T$ is set at 530~K in this
work. $f(v)$ is a correction factor to account for the transverse
velocity distribution of slowed atoms after they absorb resonant
photons in a Zeeman slower, which can be expressed as
\begin{equation}
f(v)=1-\exp\left[-\frac{r_{\rm mot}^2}{(v_rv/3)L^2/v_f^2}\right]~.
\label{CF}
\end{equation}
Here $r_{\rm mot}$ is the radius of a MOT, $L$ is the distance
between the MOT center and the end of a Zeeman slower, $v_r$
is the recoil velocity of an atom in a slowing process, and $\sqrt{v_rv/3}$ is the transverse velocity of slowed atoms~\cite{Stamper-Kurn2010}. In
Eq.~\ref{Qs}, $v_{\rm{max}}$ is the maximum velocity of entering
atoms which can be handled by a slower. For a spin-flip Zeeman
slower, $v_{\rm{max}}$ is given by
\begin{equation}
v_{\rm{max}}=\sqrt{v_{\rm{sf}}^2+2 \eta_d \cdot a_{\rm{max}} \cdot
L_d}~,\label{vMax}
\end{equation}
where $L_d$ and $\eta_d$ are the length and the actual safe factor
of the decreasing-field section, respectively. And $v_{\rm{sf}}=2 \pi \delta/k$ is the velocity of atoms which are resonant with the slowing beam at the spin-flip section, since $B$ is zero in this section. For
example, $v_{\rm{sf}}$ is 302~m/s at $\delta=-512$~MHz in our
sodium BEC apparatus.

For a given $\delta$, Eqs.~\ref{Qs}-\ref{vMax} predict that $N$
should monotonically increase with $v_{\rm max}$, i.e., $N$ increases with $\eta_d$ at a fixed $L_d$ (or $N$ increases with $L_d$ at a fixed $\eta_d$). However, our
observations shown in Fig.~\ref{decreasing}(a) are drastically different
from this prediction: at each $L_d$ studied in this paper, $N$ appears to first increase with $\eta_d$,
reach its peak $N_{\rm{max}}$ at a critical value of $\eta_d$, and
then decrease with $\eta_d$. Based on Eq.~\ref{vMax}, we can also plot
these data points as a function of $v_{\rm max}$, as shown in
Fig.~\ref{decreasing}(b). At each value of $L_d$, the agreement
between our data and a theoretical prediction derived from
Eqs.~\ref{Qs}-\ref{vMax} (dotted black line) can only
be found when $v_{\rm max}$ is smaller than $800$~m/s, which is approximately equal to the mean velocity ($\sqrt{9 \pi k_B T/8m}$~) of atoms entering our slower.

In addition, we plot $\Delta N/\Delta L_d$ as a function of
$\eta_d$ in Fig.~\ref{decreasing}(c), where $\Delta N/\Delta L_d$ 
represents the number of extra atoms in the MOT gained from elongating 
the decreasing-field section by $\Delta L_d=L_d-0.15~\rm m$. 
Figure~\ref{decreasing}(c) shows that $\Delta N/\Delta L_d$ drops to 
a much smaller value when $L_d$ is increased from 0.3~m to 0.48~m. 
This implies the ideal length of the decreasing-field 
section for our apparatus should be longer than 0.3~m and shorter than 0.48~m. It is worth noting that $\Delta N/\Delta L_d$
peaks at $\eta_{d}\sim 0.65$ for each value of $L_d$, as shown in Fig.~\ref{decreasing}(c). Interestingly, the predicted $\eta_p$ from Eq.~\ref{etap} is also 0.65 at $P=$~70~mW, and Fig.~\ref{decreasing}(a) shows $\eta_{d}\sim 0.65$ is also the position where the largest $N$ occurs. This indicates that $\eta$ of our optimized decreasing-field section is actually equal to $\eta_p$. Therefore, one important procedure in designing an optimal Zeeman
slower is as follows: 1) find out $\eta_p$ based on Eq.~\ref{etap} from the available slowing beam
intensity; 2) determine the length of the decreasing-field section from Eq.~\ref{vMax}, i.e.,
$L_d=[9\pi k_B T/8m-(2 \pi \delta /k)^2]/(2 \eta_p a_{\rm max})$; 3) find out electric currents $i$ of decreasing-field coils with our simulation program by precisely matching the simulated $B$ to a
prediction derived from Eq.~\ref{1}.
\begin{figure}[b]
\includegraphics[width=85mm]{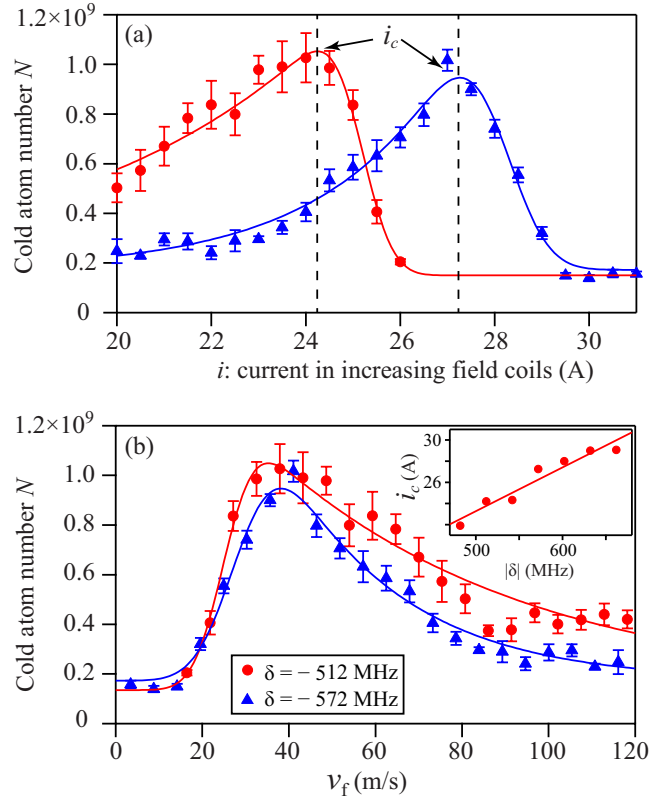}
\caption{(color online) $N$ as a
function of $i$ (a) and $v_{\rm{f}}$ (b) at $\delta=-512$~MHz (red
circles) and $\delta=-572$~MHz (blue triangles). $i_c$ at a given
$\delta$ is defined as the electric current at which $N$ peaks. Solid lines
are fits to guide the eye. Inset: $i_c$ extracted from Panel(a) as
a function of $|\delta|$. The solid line is a linear fit.}
\label{detune}
\end{figure}

\subsection{The increasing-field section}

The aforementioned discussion on the decreasing-field section applies
to the increasing-field section as well. To only study the increasing-field section, data shown in this subsection are taken at
$\eta_d=\eta_p=0.65$, $L_d=0.54~\rm m$, and $P=$~70~mW.

Our data in Fig.~\ref{detune}(a) shows that $N$ is not a monotonic function
of $i$ at a given $\delta$. $N$ first increases with $i$ and
reaches a peak at a critical value $i_{c}$, because a higher $i$
leads to a larger deceleration which means more atoms can be
slowed and captured in the MOT. When $i$ is higher than $i_{c}$,
we find that $N$ sharply decreases with $i$ due to atoms being
over-slowed. Because $v_{f}=-(2\pi \delta+\mu
B_{\rm{iMax}}/\hbar)/k$, the data points in Fig.~\ref{detune}(a)
can be converted to reveal the relationship between $N$ and $v_f$,
as shown in Fig.~\ref{detune}(b). Here $B_{\rm{iMax}}$ is the
maximum magnetic field strength created by the increasing-field
section at a fixed $i$. We find four
interesting results: $N$ always peaks around $v_f=40$~m/s $<v_c$ at
any $\delta$ within the range of $- 450~{\rm MHz} \leq \delta
\leq - 650~{\rm MHz}$; the minimal velocity of the atoms captured in the
MOT appears to be $\sim$ 20~m/s; the maximum value of $N$ does not
depend on $\delta$; and $i_c$ linearly increases with $\delta$, as shown in Fig.~\ref{detune}.
Therefore, in addition to the procedures listed in Sections 3.A and 3.B, another useful procedure in designing an
optimal spin-flip Zeeman slower is as follows: 1) choose a convenient
$\delta$, for example, $\delta$ is around $- 500~\rm MHz$ for sodium or lithium atoms; 2) find out the ideal length of the
increasing-field section from the following equation, $L_i=(v_{\rm sf}^2-v_c^2)/(2 a_s)=[(2 \pi \delta /k)^2-v_c^2]/(2
\eta_p a_{\rm max})$; 3) determine $i_c$ from a figure similar to
Fig.~\ref{detune}(a) and then set the current $i$ at a value slightly
smaller than $i_c$ in the increasing-field coils.

\subsection{The spin-flip section}
\begin{figure}[t]
\includegraphics[width=85mm]{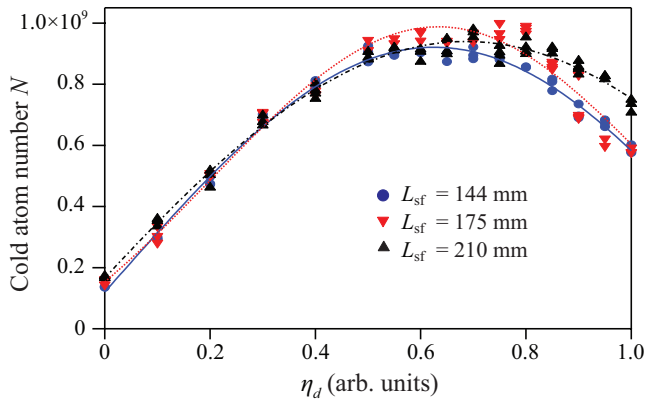}
\caption{(color online) $N$ as a function of $\eta_{d}$ at three different $L_{\rm{sf}}$, the length of the spin-flip section (see text). Lines are Gaussian fits to the
data.} \label{sf}
\end{figure}
We have also studied the contribution of the spin-flip section,
but have not found a strong correlation between the slower's efficiency and $L_{\rm{sf}}$, the length of the
spin-flip section.
Figure~\ref{sf} shows that $N$ always peaks at $\eta_d \approx \eta_p = 0.65$
for three different values of $L_{\rm{sf}}$, when $L_d$ is kept at
0.54~m, $\eta$ is set at 0.65 in the increasing-field section, $\delta$
is $- 512~\rm MHz$, and $P$ is 70~mW. This figure also implies that a
longer spin-flip section does not boost the number of atoms
captured in the MOT, as long as its length $L_{\rm{sf}}$ is
sufficient to fully re-polarize atoms. A very long $L_{\rm{sf}}$,
however, has a negative effect on the MOT capture efficiency,
because it also unavoidably reduces the solid angle of the atomic
beam when all other parameters of the system remain
unchanged.
\section{Conclusions}
We have presented the design and construction of a spin-flip
Zeeman slower controlled by a fast feedback circuit, and
demonstrated an efficient method to optimize a slower by using our
simulation program and monitoring the number of atoms trapped in
the MOT. Our data suggests that an optimal Zeeman slower may be
designed with the following procedures: 1) determine $\eta_p$ based on Eq.~\ref{etap} from the available
slowing beam intensity; 2) choose a convenient $\delta$ and find out the ideal lengths of the increasing- and
decreasing-field sections from $L_i=[(2 \pi \delta /k)^2-v_c^2]/(2
\eta_p a_{\rm max})$ and $L_d=[9\pi k_B T/8m-(2 \pi \delta /k)^2]/(2
\eta_p a_{\rm max})$, respectively; 3) set $i$ at a
value slightly smaller than $i_c$ in the increasing-field coils and
determine $i$ of decreasing-field coils with our simulation program. We have found
that a longer spin-flip section does not boost the number of atoms
captured in the MOT, as long as its length $L_{\rm{sf}}$ is
sufficient to fully re-polarize atoms. These conclusions are very
useful in designing an optimal Zeeman slower for other atomic
species, especially those with high initial velocities, for
example lithium atoms.
\section{Acknowledgments}
We thank Ian Spielman and Karl Nelson for insightful discussions,
and Jared Austin and Micah Webb for experimental assistances. We
also thank the U. S. Army Research Office, Oklahoma Center for the
Advancement of Science and Technology, and Oak Ridge Associated
Universities for financial support.


\end{document}